\documentclass[final,5p,times,twocolumn]{elsarticle}

\usepackage{amssymb}
\usepackage{lipsum}
\usepackage{xurl}
\usepackage{lineno}
\usepackage{setspace}
\usepackage[separate-uncertainty=true]{siunitx}
\usepackage{textcomp}
\usepackage{amsmath}
\usepackage{comment}
\usepackage{diagbox}
\usepackage{cellspace}
\usepackage{makecell}
\setcellgapes{4pt}
\usepackage{xcolor}
\usepackage{soul,xcolor}
\newcommand{\cirm}[1]{\textcolor{black}{#1}}

\journal{Physics Letters B}

\begin{document}
\begin{frontmatter}

\title{CPT and Lorentz symmetry tests with hydrogen using  a novel in-beam  hyperfine spectroscopy method applicable to antihydrogen experiments}

\author[CERN,SMI,UniV]{L.~Nowak}
\author[CERN]{C.~Malbrunot\corref{cor1}\fnref{label1}}
\ead{chloe.m@cern.ch}
\author[SMI]{M.~C.~Simon}
\author[SMI]{C.~Amsler}
\author[SMI]{S.~Arguedas Cuendis\fnref{label2}}
\author[SMI]{S.~Lahs\fnref{label3}}
\author[SMI,UniV]{A.~Lanz\fnref{label4}}
\author[SMI,UniV]{A.~Nanda}
\author[SMI]{M.~Wiesinger\fnref{label5}}
\author[CERN]{T.~Wolz}
%\author[SMI]{J.~Zmeskal}
\author[SMI]{E.~Widmann}

\fntext[label1]{Permanent address: TRIUMF, 4004 Wesbrook Mall, Vancouver, BC  V6T 2A3, Canada and Physics Department, McGill University, Montréal, Québec H3A 2T8, Canada and Department of Physics and Astronomy, University of British Columbia, Vancouver BC, V6T 1Z1 Canada}
\fntext[label2]{Present address: Consejo Nacional de Rectores (CONARE), San José, Geroma, 10109, Costa Rica}
\fntext[label3]{Present address:  Universit\'e Paris-Saclay, CNRS,  Laboratoire Aim\'e Cotton, 91405, Orsay, France}
\fntext[label4]{Present address:University College London, Gower St, London WC1E 6BT, United Kingdom
}
\fntext[label5]{Present address: Faculty of Physics, Ludwig-Maximilians-Universität München, Am Coulombwall 1, 85748 Garching bei München, Germany}

\affiliation[CERN]{organization={CERN},%Department and Organization
            addressline={1, Esplanade des Particules}, 
            city={Meyrin},
            postcode={CH-1211}, 
            country={Switzerland}}
 \affiliation[SMI]{
            organization={Stefan Meyer Institute for Subatomic Physics},%Department and Organization
            addressline={Dominikanerbastei 16}, 
            city={Vienna},
            postcode={A-1010}, 
            country={Austria}}
\affiliation[UniV]{organization={University of Vienna, Vienna Doctoral School in Physics},%Department and Organization
            addressline={Universitaetsring 1}, 
            city={Vienna},
            postcode={A-1010}, 
            country={Austria}}

\begin{abstract}

We present a Rabi-type measurement of two ground-state hydrogen hyperfine transitions performed in two opposite external magnetic field directions.
This puts first constraints at the level of \SI{2.3e-21}{\giga\electronvolt} on a set of coefficients of the Standard Model Extension, which were not measured by previous experiments.
Moreover, we introduce a novel method, applicable to antihydrogen hyperfine spectroscopy in a beam, that determines the zero-field hyperfine transition frequency from the two transitions measured at the same magnetic field.
Our value, $\nu_0=\SI[parse-numbers=false]{1.420\,405\,751\,63(63)}{\giga\hertz}$, is in agreement with literature at a relative precision of 0.44~ppb. This is the highest precision achieved on hydrogen in a beam, improving over previous results by a factor of 6.

\end{abstract}


\end{frontmatter}

\section{Introduction}
\label{s:intro}
Precision spectroscopy of the hydrogen atom is receiving renewed interest, in particular in the context of the proton radius puzzle~\cirm{\cite{Pohl2010, RevModPhys.94.015002}}, the determination of the fine structure constant \cirm{~\cite{Parker2018, Morel:2020tj, PhysRevLett.130.071801}}, and for comparison with its antimatter counterpart, the antihydrogen ($\overline{\text{H}}$) atom, hitherto the only synthesised stable atom made of antimatter~\cirm{\cite{BAUR1996251, Amoretti2002, Gabrielse2008, Andresen2010}}.
Techniques developed for precise measurements of hydrogen atoms are being applied to antihydrogen ~\cirm{\cite{Ahmadi:2018vo, Baker:2021us}} and new techniques are tested on hydrogen ~\cirm{ \cite{10.1063/5.0070037, Azevedo:2023te, Jones_2022, Fujiwara2020}} before being deployed in the more challenging environment of antihydrogen experiments.
One of the main motivations for matter/antimatter comparisons is to test CPT symmetry (the combination of the three discrete symmetries: Charge conjugation, Parity, and Time reversal).
Comparing the sensitivities of different tests of CPT symmetry requires the use of a unified framework to parameterise deviations from this symmetry.
A framework that serves this purpose is the Standard Model Extension (SME)~\cite{PhysRevD.55.6760, PhysRevD.58.116002} which is an effective field theory that generalises the Standard Model (SM) Lagrangian by adding terms violating CPT and Lorentz symmetry.
The models for testing Lorentz and CPT symmetry in atomic spectroscopy experiments~\cite{PhysRevLett.82.2254,PhysRevD.92.056002} reveal that the sensitivity to coefficients that quantify a breaking of CPT symmetry, known as SME coefficients, depends on the absolute precision of the experiment. 
From this viewpoint the hyperfine splitting could provide the most stringent test, as this transition has been determined with unrivalled absolute uncertainties of a few mHz using hydrogen masers~\cite{hellwig1970measurement, essen1971frequency, essen1973hydrogen, KARSHENBOIM20051, ramsey1990experiments, RamseyQE:1190}.
Such precise maser measurements have provided tight constraints on SME coefficients accessible through searches for sidereal variation~\cite{PhysRevD.63.111101, PhysRevA.68.063807} within the so-called \textit{minimal}~SME~\cite{PhysRevLett.82.2254}, restricted to Lorentz-violating operator with mass dimension $d\leq4$, expected to contain the dominant effects at low energies.
An extension of the framework was since developed, including operators of arbitrary dimensions~\cite{PhysRevD.88.096006, PhysRevD.92.056002}.
Commonly appearing combinations of SME coefficients are defined as effective SME coefficients.
In this work we consider the non-relativistic (NR) effective coefficients~\cite{PhysRevD.88.096006}.
We denote the NR coefficients in a laboratory frame on the Earth as $\mathcal{K}^{\textrm{NR,lab}}$.
The relation to the NR coefficients in the Sun-centered frame, the frame commonly used to report the limits on the SME coefficients, with the ones in the laboratory frame is given by the transformation:
\begin{equation}
    \label{eq:Klabagain}
    \begin{split}
    \mathcal{K}^{\textrm{NR,lab}}_{w_{k10}} =  {\mathcal{K}^{\textrm{NR,Sun}}_{w_{k10}}\cos{\theta}}  - \sqrt{2}\Re\mathcal{K}^{\textrm{NR,Sun}}_{w_{k11}} \sin{\theta} \cos{\omega_{\oplus}T_{\oplus}}\\+
    \sqrt{2}\Im\mathcal{K}^{\textrm{NR,Sun}}_{w_{k11}} \sin{\theta} \sin{\omega_{\oplus}T_{\oplus}}
    \end{split}
\end{equation}

where $w$ stands for electron or proton, $k$ represents the power of the fermion's momentum of the operator coupled to the coefficient,
\cirm{
$\theta$ is the angle between the Earth's rotation axis and the experimental static magnetic field aligning the atoms,
}
and $T_{\oplus}$ represents the sidereal time, which is a location-dependent time that is a convenient offset of the time coordinate of the Sun-centered frame.
Searches for variations of $\mathcal{K}^{\textrm{NR,lab}}_{w_{k10}}$ at the sidereal frequency $\omega_{\oplus}$  using a hydrogen maser have constrained the real and imaginary parts of $\mathcal{K}^{\textrm{NR,Sun}}_{w_{k11}}$ while $\mathcal{K}^{\textrm{NR,Sun}}_{w_{k10}}$ remained unconstrained.

\begin{figure}[ht!]
    \centering 
    \includegraphics[width=0.5\textwidth]{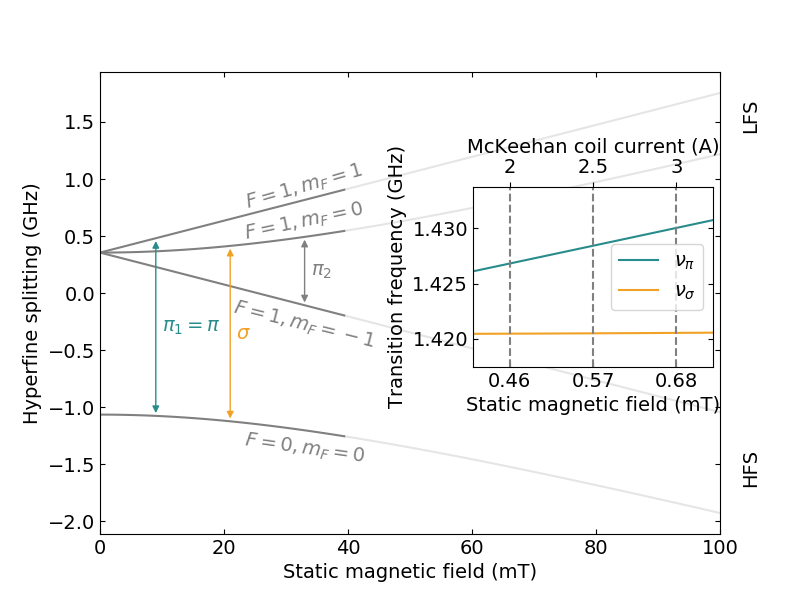}	
    \caption{
    Breit-Rabi diagram: dependence of ground-state hydrogen hyperfine energy states on an external static magnetic field ($B_\text{stat}$).
    The energy levels that increase/decrease towards higher magnetic field amplitude are labelled low/high field seeking (LFS/HFS) states.
    The two states with $m_F = \pm 1$ experience a strong linear Zeeman shift.
    The mixing of the two states with $m_F$=0 results in a weaker, second order Zeeman shift and also in a suppressed sensitivity to potential CPT or Lorentz symmetry violations.
    The inset illustrates the drastically different field dependence for the $\sigma$ and $\pi_1$ transition within the region of magnetic fields probed in this work.
    } 
    \label{fig_BreitRabi}%
\end{figure}

This precise maser technique unfortunately cannot be applied to $\overline{\textrm{H}}$ hyperfine spectroscopy as it relies on storage in a bulb.
One approach is to work with magnetically trapped $\overline{\textrm{H}}$ as done successfully by the ALPHA collaboration \cite{Ahmadi:2017wxa}.
First hyperfine spectroscopy results have been reported \cite{Ahmadi:2017} although control of systematic uncertainties in the highly inhomogeneous magnetic field environment poses a challenge.
An alternative approach for $\overline{\textrm{H}}$ hyperfine spectroscopy has been proposed by the ASACUSA collaboration~\cite{WidmannEtAl2001, Widmann200431}.
Producing a beam of $\overline{\textrm{H}}$ \cite{Kuroda:2014uxa} enables Rabi-type measurements~\cite{PhysRev.53.318, PhysRev.55.526}, in which the hyperfine interaction takes place in a region of well-controlled fields outside of the $\overline{\textrm{H}}$ production trap.
ASACUSA tested its $\overline{\textrm{H}}$ Rabi-spectrometer~\cite{MALBRUNOT2019110} with hydrogen and reached an absolute precision of \SI{3.8}{\hertz} (corresponding to a relative precision of \SI{2.7}{ppb}) on the hyperfine splitting frequency~\cite{diermaier2017beam}, thereby improving over previous in-beam measurements~\cite{PhysRev.88.184, PhysRev.100.1188}.
In that work, the $\sigma$ transition ($F,m_F$:1,0$\rightarrow$0,0, where $F$ represents the total angular momentum quantum number and $m_F$ its $z$-axis projection, see Fig.~\ref{fig_BreitRabi}) was measured at various magnetic field values and extrapolated to zero field.
However, the sensitivity of the $\sigma$ transition to SME effects is strongly suppressed at low fields and eventually vanishes at zero field.
Similarly, when the difference of the $\pi_1$ ($F,m_{F}$: 1,1 $\rightarrow$ 0,0) and $\pi_2$ ($F,m_{F}$: 1,0 $\rightarrow$ 1,-1) transitions is used to extract the zero-field hyperfine splitting~\cite{Ahmadi:2017} one becomes insensitive to SME coefficients.

Here we employ quasi-simultaneous measurements of the $\sigma$ and $\pi_1$ transitions (hereafter referred to as $\pi$ transition) and use the results in two ways:
by measuring the $\pi$ transition at opposite magnetic field orientation we constrain coefficients embedded in $\mathcal{K}^{\textrm{NR,Sun}}_{w_{k10}}$. 
Using two transitions enables to disentangle SME effects from shifts originating from magnetic field uncertainties imposed by the change of field orientation.
Furthermore, for each resonance pair the zero-field hyperfine splitting can be calculated.
This novel method is applicable to antihydrogen as an accurate result is already obtained from the measurement of a single ($\sigma$,~$\pi$) transition pair, while extrapolation to zero field requires multiple measurements.

\section{Experimental setup}
\label{s:exp_setup}

\begin{figure*}[ht!]
	\centering 
	\includegraphics[width=1\textwidth]{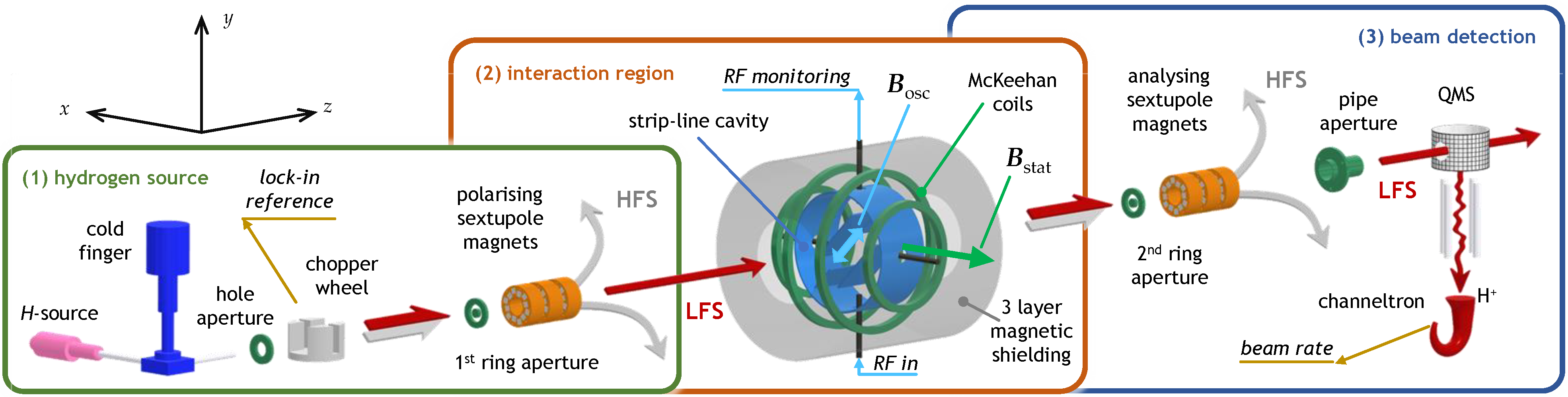}	
	\caption{
    Schematics of the experimental apparatus (total length $\sim\SI{5}{m}$, not to scale).
    The hydrogen source region (1) consists of a microwave-driven plasma dissociating H$_2$ molecules, a cryostat cooling hydrogen atoms passing through a PTFE tubing, a chopper modulating the beam, and sextupole magnets for beam polarisation.
    In the interaction region (2) a strip-line cavity provides the oscillating magnetic field \textbf{$B_\text{osc}$} at an angle of 45$^{\circ}$ with respect to the static external magnetic field \textbf{$B_\text{stat}$} produced by a set of McKeehan coils.
    The assembly is enclosed in a three-layer cylindrical mu-metal shielding. The beam is then spin-analysed in the detection region (3) by sextupole magnets before it enters a quadrupole mass spectrometer that selectively counts protons.}  
	\label{fig_Setup}%
\end{figure*}

The measurement principle follows Rabi's magnetic resonance spectroscopy \cite{PhysRev.53.318}:
polarised atoms (low field seekers - LFS) undergo a quantum transition in the presence of an oscillating magnetic field.
A magnetic field gradient then separates the states and focuses LFS atoms on a detector.
The count rate at the detector is acquired as a function of the frequency of the oscillating field to reveal a resonance pattern from which the transition frequency can be determined.

The apparatus employed here is a modified version of the beamline designed to characterise the ASACUSA antihydrogen hyperfine spectrometer~\cite{MALBRUNOT2019110}, which yielded the previous best in-beam measurement of the hydrogen hyperfine structure~\cite{diermaier2017beam}. 
Figure~\ref{fig_Setup} shows a sketch of the experimental setup.
It consists of a source of cold, modulated, and polarised atomic hydrogen (1), a hyperfine interaction region providing well-controlled static and oscillating magnetic fields (2), and a detection system based on state selection and single ion counting after electron impact ionisation (3).

Hydrogen molecules are dissociated in the microwave-driven plasma of an Evenson cavity~\cite{Evenson:1965}.
The hydrogen atoms lose energy through collisions in a PTFE tubing sandwiched between two aluminum blocks connected to a cold finger stabilised at \SI{27}{\kelvin}.
The cooled hydrogen beam then passes through an aperture to a second vacuum chamber housing a chopper which modulates the beam at a frequency of \SI{66}{\hertz} for lock-in amplification allowing for background suppression against residual hydrogen in the detection region.
At the exit of this vacuum chamber a ring aperture ($1^{\textrm{st}}$ ring aperture) blocks the central part of the divergent beam before it enters the magnetic sextupole field produced by permanent magnets arranged in a circular Halbach array~\cite{wiesinger:2017}.
At the typical radius of the annular beam of \SI{15}{\milli\meter} the sextupole field strength reaches \SI{190}{\milli\tesla}.
At such high fields the two LFS states have comparable magnetic moments, thus the trajectory restriction by the ring apertures selects similar velocities for the two initial states.

\begin{figure}[ht]
	\centering 
	\includegraphics[width=0.5\textwidth]{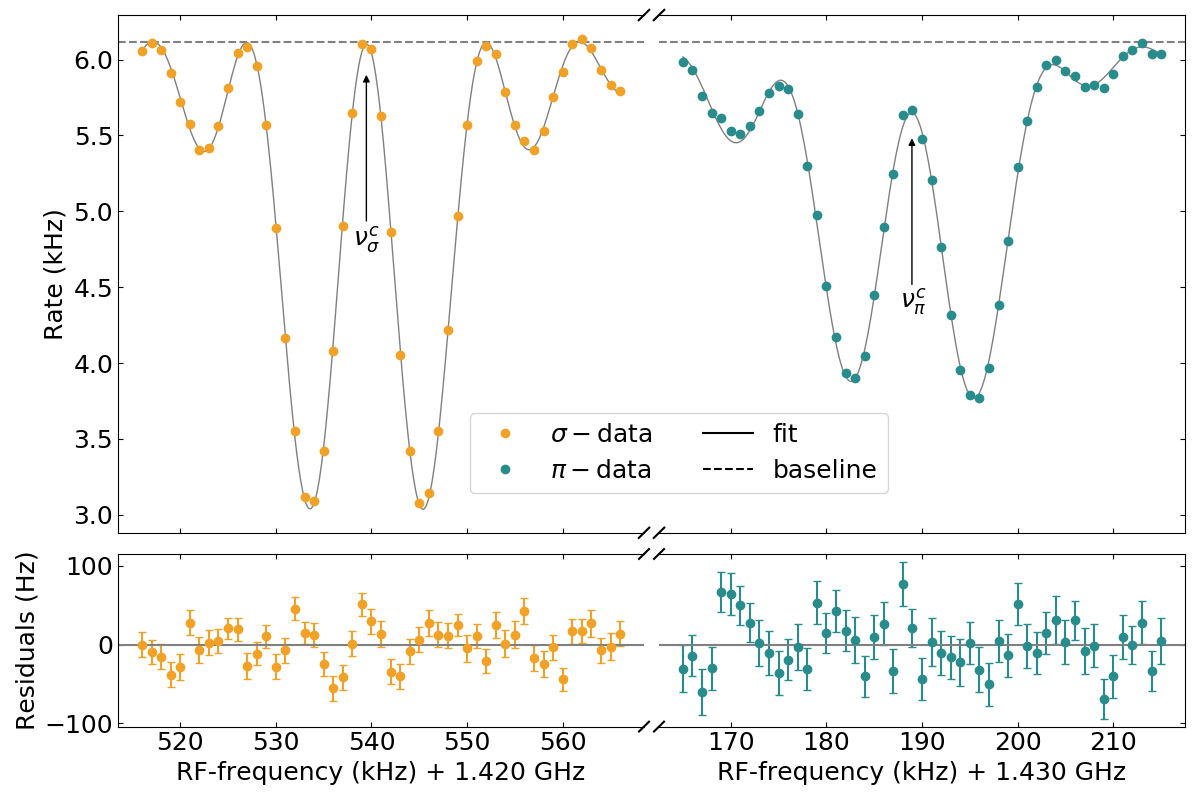}	
	\caption{
    Example of subsequently recorded $\sigma$ (left) and $\pi$ (right) resonances.
    The effect of the magnetic field inhomogeneity on the resonance shape is clearly visible on the $\pi$ resonance.
    In particular, the resonance is asymmetric and at $\nu^c_{\pi}$ the rate is not returning to the baseline as for the $\sigma$ resonance.
    This lineshape is the result of different field inhomogeneities and is very well accounted for by the fitting algorithm as shown from the residuals. The dotted line indicates the baseline $R_0$ which is a fit parameter.
    The field inhomogeneities are also responsible for what appears to be different beam rate drops between the $\sigma$ and $\pi$ transitions.
    In fact the two LFS states ($F=1, m_F=0,1$) are about equally populated. 
    } 
	\label{fig_Res}
\end{figure}

The polarised beam then enters the radio-frequency (RF) cavity~\cite{MALBRUNOT2019110}, which was designed with a large geometrical acceptance (open diameter of \SI{100}{\milli\meter}) as a compensation measure for the small yield of $\overline{\textrm{H}}$ experiments. 
At the entrance and exit of the cavity, the RF field is contained by two meshes with a combined transparency of 95\% and separated by \SI{105.5}{\milli\meter}, i.e. half a wavelength of the hyperfine transition.
A signal generator coupled to an amplifier provides the RF waves, which are guided to an antenna in the cavity via a coupling-maximising double stub tuner and a vacuum feedthrough.
A second antenna picks up the RF power for monitoring by a spectrum analyser.
Both, the signal generator and spectrum analyser are frequency-stabilised by a GPS-disciplined rubidium clock with a long-term relative frequency stability on the order of $\sim1\times10^{-12}$.
A standing wave forms in the cavity leading to a sinusoidal variation of the oscillating magnetic field, $B_\text{osc}$, in the beam propagation direction, causing a double-dip resonance lineshape, as illustrated in Fig.~\ref{fig_Res}.
For the derivation of the resonance shape see Ref.~\cite{diermaier2017beam}.
The microwave cavity is tilted by 45$^{\circ}$ compared to the previous arrangement. Therefore  \textbf{$B_\text{osc}$} has a component parallel and perpendicular to the static magnetic field~\cite{doi:10.1142/9789811213984_0001} as needed to drive the $\sigma$ and $\pi$ transition, respectively.

As shown in the inset of Fig.~\ref{fig_BreitRabi}, the $\pi$ transition is much more sensitive to the magnetic field and thus to magnetic field inhomogeneities.
Hence, the Helmholtz coils of the previous setup were replaced by a McKeehan configuration~\cite{McKeehan}, consisting of two pairs of coils of different radii~\cite{cuendis:2017}, powered by a precision bipolar current supply.
The cuboidal two-layer magnetic mu-metal shielding was upgraded to a three-layer cylindrical shielding.
Overall, the homogeneity of the static magnetic field, $B_{\text{stat}}$, within the cavity volume was improved by a factor $\sim 20$, resulting in $\sigma_{ B_{\text{stat}}}/B_{\text{stat}}\sim\SI{0.05}{\%}$ as established from fluxgate measurements \cite{cuendis:2017}.

The atoms in LFS states entering the cavity undergo transitions when the RF frequency and power are in the appropriate regime.
Therefore the beam downstream of the cavity will contain a fraction of the ($F=0$) high field seeking (HFS) states again, that can be separated from the LFS states by another magnetic field gradient. 
The large acceptance (\SI{100}{\milli\meter} opening diameter) superconducting sextupole magnet employed for this state analysis in~\cite{diermaier2017beam}, now in use in the ASACUSA $\overline{\textrm{H}}$ experiment, was replaced by permanent sextupole magnets equivalent to the ones providing the initial beam polarisation.
For the hydrogen experiment an open diameter of \SI{40}{\milli\meter} is sufficient, and the required field gradient can be achieved with a smaller pole field strength. 
These analysing magnets remove all atoms that transitioned to HFS states and focus the remaining LFS atoms through a pipe aperture into a quadrupole mass spectrometer (QMS) located in the final detection chamber.
The QMS ionises the hydrogen atoms by electron impact.
Protons are selectively guided to a channeltron for counting at typical rates of $\sim5-\SI{10}{\kilo\hertz}$ (see Fig.~\ref{fig_Res}).
The pressures in the five differentially pumped vacuum chambers reduce from about \SI[print-unity-mantissa=false]{e-3}{\pascal} in the first chamber by roughly one order of magnitude after each of the four apertures to about \SI[print-unity-mantissa=false]{e-8}{\pascal} in the detection chamber.

\section{Measurement}

Probing $\mathcal{K}^{\textrm{NR,Sun}}_{w_{k10}}$ requires a change of the component of the static magnetic field, that projects onto the Earth's rotation axis.
An efficient way to do this is by changing the direction of the field.
This corresponds to changing the polarity of the current in the coils, therefore the two directions are labelled positive ($+$) and negative ($-$).

SME effects on the $\sigma$ transition are suppressed by roughly two orders of magnitude with respect to the $\pi$ transition (see Eq.~\ref{eq:SME_sigma} and Eq.~\ref{eq:SME_pi} and associated discussion).
We hence use the $\sigma$ transition to provide a concomitant measurement of the magnetic field inside the interaction region via the following method:
based on the magnetic field dependence of the transitions (deduced from the Breit-Rabi formulae \cite{PhysRev.38.2082.2}, see Eq.~\ref{eq:BreitRabi:sigma} and Eq.~\ref{eq:BreitRabi:Pi} below) we calculate from the extracted centre frequency of the $\sigma$ transition ($\nu^c_{\sigma}$, see Fig.~\ref{fig_Res}) the expected $\pi$ transition frequency ($\nu^c_{\pi\leftarrow\sigma}$) at the same magnetic field and compare it to the extracted centre frequency from our $\pi$ measurement ($\nu^c_{\pi}$).
By comparing the obtained difference $\Delta \nu_{\pi} = \nu^c_\pi- \nu^c_{\pi\leftarrow\sigma}$ for negative and positive polarities most systematic effects cancel, while potential SME shifts are enhanced.
This double difference ($\Delta \nu_\pi^+ \! - \!  \Delta \nu_\pi^-$) is the final quantity used to constrain SME effects.
The impact of extracting  $\nu^c_{\pi\leftarrow\sigma}$ from $\nu^c_{\sigma}$  on the sensitivity to different SME coefficients will be discussed in section~\ref{s:Results}.

It is worth noting that the second order Zeeman shift of the $\sigma$ transition has two consequences.
On the one hand, a small frequency uncertainty translates into a relatively large field uncertainty.
The acquisition of more data for $\sigma$ (\SI{45}{\minute} per resonance) than for $\pi$ (\SI{15}{\minute}) transitions mitigates this effect.
On the other hand, with the same absolute frequency uncertainty, an increased precision on the magnetic field is reached at higher fields.
Therefore, measurements were taken at relatively high fields  up to \SI{0.68}{\milli\tesla}, corresponding to a current $I_c=\SI{3}{\ampere}$ on the McKeehan coils.

A measurement series consisted of successive $\sigma$ and $\pi$ transitions acquisitions at negative and positive polarities (a total of four transitions were acquired within \SI{2}{\hour}).
For each transition the oscillating field frequency was varied in a random sequence of 51 points over a range of \SI{50}{\kilo\hertz} (step size of \SI{1}{\kilo\hertz}).
The absolute frequency was shifted by software before analysis, thus effectively blinding the data.
Figure~\ref{fig_Res} shows an example of the resonances obtained at \SI{-3}{\ampere}.
Prior to acquiring rate spectra in the frequency domain the RF input power was scanned to observe Rabi oscillations at the frequency $\Omega_\text{Rabi} \propto B_\text{osc}$.
The power in the cavity was then set to drive the first state population inversion (called $\pi$-pulse in conventional Rabi spectroscopy) as indicated in Fig.~\ref{fig_2D}, where $\nu^c$ is best resolved.
Over the course of two months a total of 576 resonance pairs were recorded at six different coil current settings ($I_c=\pm$\SI{2}{\ampere}, $\pm$\SI{2.5}{\ampere}, and $\pm$\SI{3}{\ampere}) with 39 pairs not passing data quality selection due to high rate instabilities.

In addition to the acquisition of ($\sigma$,~$\pi$) resonance pairs, a series of measurements of the $\pi$ transitions were taken at different oscillating magnetic field strengths, resulting in two-dimensional maps, as exemplified in Fig.~\ref{fig_2D}.
For these $\pi$ state-conversion data maps, recorded at the same six different current settings as the main data, the oscillating field frequency was stepped in a random sequence of 41 points over a range of \SI{80}{\kilo\hertz} (step size of \SI{2}{\kilo\hertz}) and the oscillating magnetic field was scanned linearly up to values allowing to observe the third state population inversion.
These two-dimensional $\pi$ maps provide an independent data set to assess the static magnetic field quality in the cavity volume.

\section{Analysis}

\begin{figure}[ht]
	\centering 
	\includegraphics[width=0.5\textwidth]{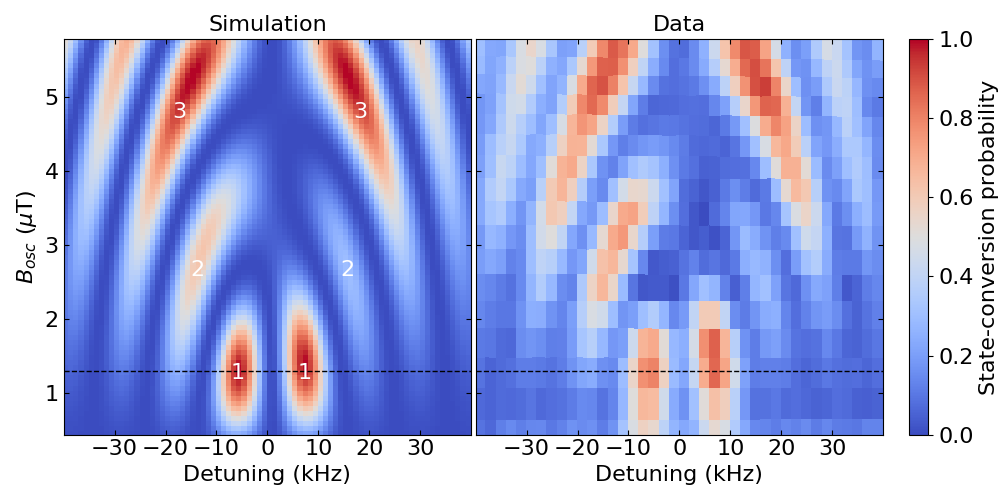}	
	\caption{
    Comparison of simulated and measured probabilities of conversion from LFS to HFS states for the $\pi$ transition at \SI{-3}{\ampere} as a function of the detuning ($\nu_{\text{RF}}-\nu_{\pi}$) and the amplitude of the oscillating magnetic field $B_\text{osc}$.
    The simulated map results from numerical solutions of the von Neumann equation for the hydrogen four-level hyperfine system considering the oscillating and static magnetic fields configuration of the experiment.
    The black dashed horizontal line indicates the typical RF field driving strength used for the acquisition of ($\sigma$,~$\pi$) pairs.
    The effect of the non-homogeneous static magnetic field causing an asymmetry in the resonance can clearly be seen and is particularly strong in the region of the second state population inversion (labelled 2). This general feature is very well reproduced in the simulation. 
    } 
	\label{fig_2D}
\end{figure}

The extraction of the centre frequencies $\nu^\text{c}$ from the data builds upon the fit procedure developed for the experiment addressing only $\sigma$ transitions~\cite{diermaier2017beam}.
The rate fit function $\mathcal{R} (\nu_{\text{RF}};R_0,p,\cirm{\nu^{c}},\textrm{v}_{\textrm{H}},\Delta\nu^{x,y},B_\text{osc},\Delta B^z_1,\Delta B^z_2)$ employed here has the frequency $\nu_{\text{RF}}$ as variable and eight parameters:
the first two parameters, the baseline $R_0$ and the polarisation $p$ scale the state conversion probability to the observed rate.
The next four parameters enter in the probability function constructed from a spline interpolation on simulated two-dimensional maps (see Fig.~\ref{fig_2D} left).
The centre frequency $\nu^c$ is thus obtained as well as the velocity of the atoms $\textrm{v}_{\textrm{H}}$,  the frequency broadening $\Delta\nu^{x,y}$ due to inhomogeneities of the static magnetic fields in the (x,y)-plane, and the oscillating magnetic field strength $B_\text{osc}$.
$\textrm{v}_{\textrm{H}}$ and $\Delta\nu^{x,y}$ were extracted from all resonances and then fixed in the final fit to an averaged value.
Two different velocity values, of \SI{978}{\meter\per\second} and \SI{1054}{\meter\per\second}, were obtained for $\sigma$ and $\pi$ transitions, respectively.
The velocity difference originates from the two involved initial states, which transmit sightly differently through the magnetic sextupole fields.
\cirm{The trajectory selection through the apertures also reduced the velocity spread in comparison to the experiment addressing only $\sigma$ transitions~\cite{diermaier2017beam} making it a negligible parameter in this work.}
$\Delta\nu^{x,y}$ additionally depends on the current $I_c$ and was hence fixed for each corresponding subset of data.

Finally, the parameters $\Delta B^z_1$ and $\Delta B^z_2$ encode the first and second-order deviations from a uniform  $B_{\textrm{stat}}$ in the beam propagation direction (z) and  modify the probability map underlying the spline interpolations.
Magnetic field measurements taken with fluxgates inside the cavity, coils and shields assembly prior to their installation in the hydrogen beamline revealed those field inhomogeneities.
The magnitude of the field, with the dominant component originating from the McKeehan coils aligned along the $x$-direction, is well approximated by a second order polynomial:
\begin{equation}
    \label{eq:staticBfield}
    \centering 
    B_{\textrm{stat}}(z) =
    B_{0} + \Delta B^z_1 \ z +\Delta B^z_2(\ z^2-L_{\textrm{cav}}^2/12),
\end{equation}
where $z=0$ is the centre of the RF cavity of length $L_{\textrm{cav}}$, and
$B_0 = \overline{B_{\text{stat}}}(z \in [-L_{\text{cav}}/2,L_{\text{cav}}/2])$
is the average magnetic field determining the centre frequency of the transitions: $\nu^c_{\sigma , \pi} = \nu_{\sigma , \pi} (B_0)$.
The effects of the $z$-dependence of $B_{\text{stat}}$ have to be taken into account by using different probability maps.
Sets for various values of $\Delta B^z_1$ and $\Delta B^z_2$ have been produced by solving the von Neumann equation of the four-level system, where the spatial dependence of $B_{\text{stat}}$ translates into a time dependence via the atoms velocity.
The optimal values were obtained by calculating the least-square deviations of such maps to the measured $\pi$ probability maps (see Fig.~\ref{fig_2D} right).
The quadratic coefficient $\Delta B^z_2$ turned out to be crucial to explain asymmetries in the line profiles.
Those are strongest in the region of the second state population inversion. 
Hence the $\pi$ data maps facilitate an accurate determination of $\Delta B^z_2$.
In contrast $\Delta B^z_1$ had much weaker effects and an accurate determination was hampered by correlations with other fit parameters.
Therefore the linear parameter was set to zero and the systematic effect resulting from this choice was assessed in the same way as for the other two fixed fit parameters:
the uncertainties were conservatively estimated by varying their value around the central value by 1 or 3 standard deviations.
The resulting changes in $\nu^c$ were taken as the uncertainty.
Note that the uncertainty of the parameter $\Delta B^z_1$ is asymmetric as positive and negative deviations from the chosen value of zero cause negative shifts of $\nu^\text{c}$.

\section{Results}
\label{s:Results}

\begin{figure}[ht!]
	\centering 
	\includegraphics[width=0.4\textwidth]{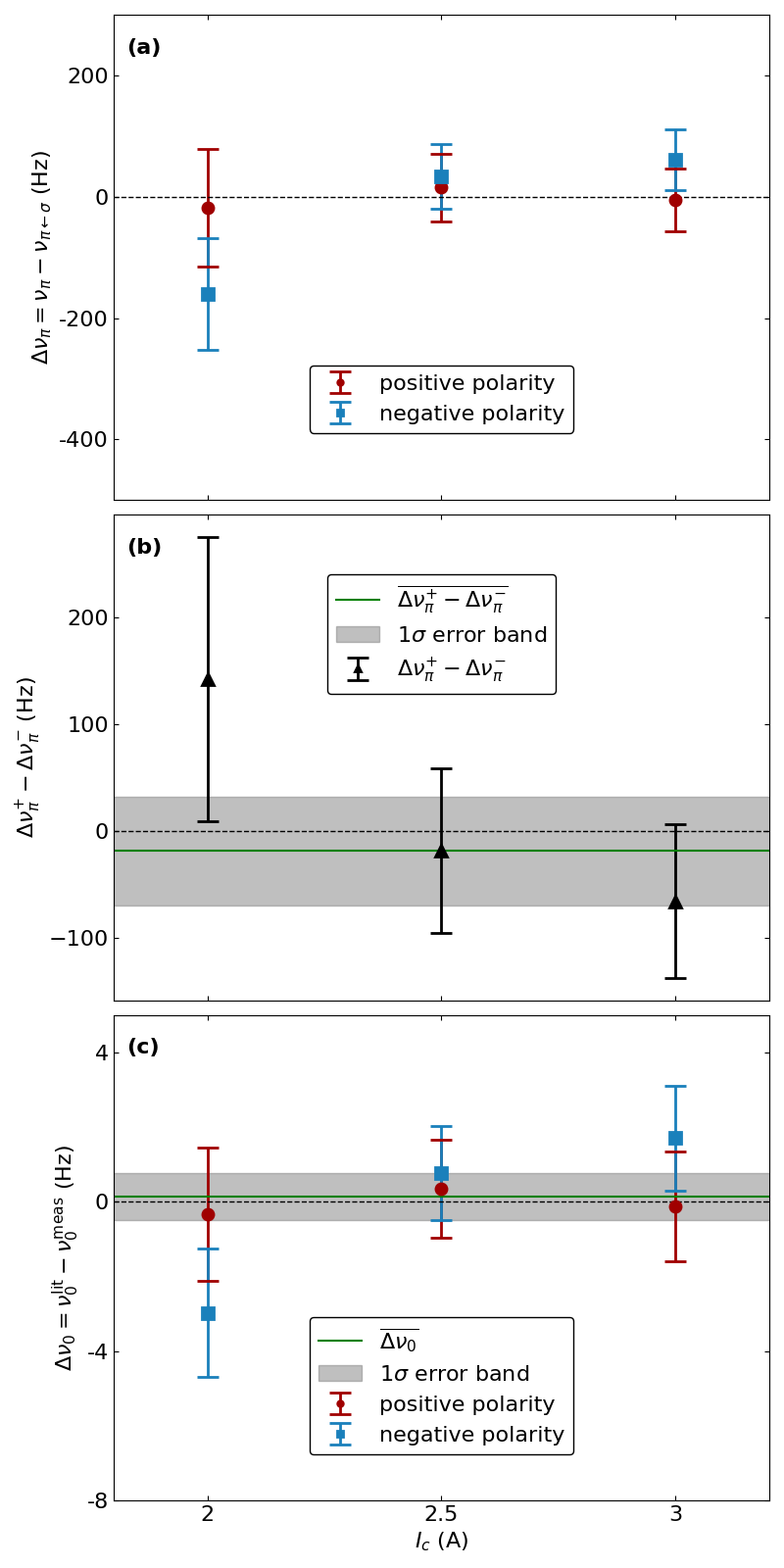}	
	\caption{
    Experimental results at the three McKeehan coil currents.
    a) Difference between the measured $\pi$ transition frequency ($\nu_{\pi}$) and the expected value ($\nu_{\pi\leftarrow\sigma}$) calculated from the Breit-Rabi formula based on the value of the magnetic field determined by the $\sigma$ transition.
    This difference is provided for positive ($\Delta\nu_{\pi}^+$) and negative polarities ($\Delta\nu_{\pi}^-$) and is consistent with expectation for both polarities.
    b) The consistency between the negative and positive polarity measurements provides a constraint on the magnitude of the coefficient of some SME terms (see text).
    The average of the difference between positive and negative polarities is consistent with zero at a precision of $\SI{51}{\hertz}$ indicated by the grey band.
    c) The quasi-simultaneous measurements of $\pi$ and $\sigma$ transitions at the same magnetic field enable the determination of the zero-field hyperfine splitting $\nu_0^\text{meas}$.
    The average is consistent with literature ($\nu_0^{\textrm{lit}}$) with an absolute precision of \SI{0.63}{\hertz}.
}
	\label{fig_Results}%
\end{figure}

The unblinded results are summarised in Fig.~\ref{fig_Results}. 
The centre frequency \cirm{$\nu^c_{\sigma}$} extracted for each $\sigma$ transition in a ($\sigma$,~$\pi$) data pair is used to compute, via the Breit-Rabi formula, the expected $\nu^c_{\pi\leftarrow\sigma}$ (using the literature value $\nu_0^{lit}=\SI[parse-numbers=false]{1.420\,405\,751\,768(1)}{\giga\hertz}$ \cite{KARSHENBOIM20051}).
This value is then compared to $\nu^c_{\pi}$ extracted from the $\pi$ resonance, see Fig.~\ref{fig_Results}~(a).
The difference $\Delta\nu_{\pi}^+ \! - \! \Delta\nu_{\pi}^-$ between the two values at the two opposite magnetic field orientations constitutes a test of the SME, see Fig.~\ref{fig_Results}~(b).
We note that the second-order corrections to the Breit-Rabi formula computed in Ref.~\cite{PhysRevA.73.052506, Shabaev} have, at the low fields used in the experiment, negligible effect on this result as well as on the extracted zero-field hyperfine splitting value, and are thus not considered.
The statistical uncertainty on $\Delta\nu_{\pi}^+ \! - \! \Delta\nu_{\pi}^-$ is primarily dominated by the propagation of the $\nu^c_{\sigma}$ uncertainty due to the second order dependency of the $\sigma$ transition on the magnetic field.
Table~\ref{table:errorbudget_stat_sys_rearranged} indicates the scaling factor $\partial \nu_\pi / \partial \nu_\sigma$ for every current which, for instance at \SI{+2}{\ampere}, leads to a $\sim$\SI{96}{\hertz} uncertainty on $\nu_{\pi\leftarrow\sigma}$ from an initial statistical uncertainty of \SI{1.75}{\hertz} on $\nu_\sigma$.
The correlation in the systematic uncertainties at different currents was taken into account following Ref.~\cite{Erler2015,BARLOW2021164864}.
Systematic uncertainties resulting from the choice of $\Delta B^z_2$ and $\Delta B^z_1$ values and from the fixation of the parameters $v_\text{H}$ and $\Delta\nu^{x,y}$ were added in quadrature and account for less than $\sim\SI{5}{\hertz}$, see Table~\ref{table:errorbudget_stat_sys_rearranged}.
A $\sqrt{}(\chi_\text{red}^2)$ inflation was adopted to take into account the quality of the individual lineshape fits and contributed to the statistical uncertainty. 

\begin{table}[ht]%[H]
\centering
 \makegapedcells\begin{tabular}{|c|*{6}{c|}}
\hline
$I_\text{c}$ & \multicolumn{2}{c|}{\SI{2.0}{\ampere}}  & \multicolumn{2}{c|}{\SI{2.5}{\ampere}}  & \multicolumn{2}{c|}{\SI{3.0}{\ampere}} \\
$B_\text{stat}$ & \multicolumn{2}{c|}{\SI{0.46}{\milli \tesla}}  & \multicolumn{2}{c|}{\SI{0.57}{\milli \tesla}}  & \multicolumn{2}{c|}{\SI{0.68}{\milli \tesla}} \\
$\partial \nu_\pi / \partial \nu_\sigma$ & \multicolumn{2}{c|}{55.0}  & \multicolumn{2}{c|}{44.1}  & \multicolumn{2}{c|}{36.8} \\
\# pairs $+/-$& \multicolumn{2}{c|}{56 \ / \ 55} & \multicolumn{2}{c|}{121 \ / \ 120} & \multicolumn{2}{c|}{93 \ / \ 92} \\
\hline
errors (Hz) & stat. & sys. & stat. & sys. & stat. & sys. \\
\hline
\hline
$\nu_\sigma^+$ & 1.75 & 0.14 & 1.26 & 0.20 & 1.40 & 0.22 \\
$\nu_\pi^+$    & 2.37 & $^{+3.19}_{-4.25}$ & 1.69 & $^{+3.25}_{-4.30}$ & 2.11 & $^{+3.26}_{-4.31}$ \\
\hline
$\nu_{\pi \leftarrow \sigma}^+$ & 96.3 & 7.70 & 55.7 & 8.82 & 51.7 & 8.10 \\
$\Delta\nu_\pi^+$ & 96.4 & $^{+8.33}_{-8.79}$ & 55.8 & $^{+9.40}_{-9.81}$ & 51.8 & $^{+8.73}_{-9.18}$ \\
\hline
\hline
$\nu_\sigma^-$ & 1.67 & 0.19 & 1.21 & 0.23 & 1.35 & 0.21 \\
$\nu_\pi^-$ & 2.93 & $^{+3.18}_{-3.66}$ & 2.19 & $^{+3.34}_{-3.80}$ & 2.69 & $^{+3.38}_{-3.84}$ \\
\hline
$\nu_{\pi \leftarrow \sigma}^-$ & 91.9 & 10.5 & 53.1 & 10.1 & 49.7 & 7.73 \\
$\Delta\nu_\pi^-$ & 92.0 & $^{+10.9}_{-11.1}$ & 53.2 & $^{+10.7}_{-10.8}$ & 49.8 & $^{+8.44}_{-8.64}$ \\
\hline
\hline
$\Delta \nu_\pi^+ - \Delta \nu_\pi^-$ & 133 & $^{+13.7}_{-14.1}$ & 77.1 & $^{+14.2}_{-14.6}$ & 71.8 & $^{+12.2}_{-12.6}$ \\
\hline
\end{tabular}
\caption{
    Error budget (in Hz).
    Systematic effects have been investigated for parameters that were fixed during the final fit procedure, namely $\textrm{v}_{\textrm{H}}$, $\Delta_\nu^{x,y}$, and both $\Delta B^z$ coefficients and added in quadrature.
    Asymmetric errors originate from the one-sided error of $\Delta B^z_1$ (negligible for $\sigma$ transitions and in the final result after rounding to two significant digits).
    The statistical errors are slightly larger for $\pi$ measurements due to the three times shorter acquisition time compared to $\sigma$ measurements and the altered lineshape.
    In the case of $\nu_{\pi \leftarrow \sigma}$ the field-dependent uncertainty propagation factor $\partial \nu_\pi / \partial \nu_\sigma$ (given in the table header) has to be taken into account.
    The uncertainty on $\Delta \nu_\pi$ is obtained by adding in quadrature those of $\nu_\pi$ and $\nu_{\pi \leftarrow \sigma}$, and is clearly dominated by the latter.
    Finally, the difference between $\Delta \nu_\pi$ for positive and negative polarity is formed where the uncertainties are again added in quadrature.
    Sample sizes for ($\sigma$,~$\pi$) pairs for each current and polarity are listed in the table header.
\label{table:errorbudget_stat_sys_rearranged}
}
\end{table}

The result, $\Delta \nu_{\pi}^+ \! - \! \Delta\nu_{\pi}^-=\SI{- 19 \pm 51}{\hertz}$ is consistent with the SM and allows to constrain the SME Sun-centred frame coefficients \cirm{$\mathcal{K}^{\textrm{NR,Sun}}_{w_{kjm}}$. The coefficients that contribute to the relevant energy shift are limited to coefficients with $j \le 1$. The contributing ones with $j=0$ are the isotropic spin-independent coefficients $\mathcal{V}^{\textrm{NR,Sun}}_{w_{k00}}$ that shift all ground-state hyperfine sub-level by the same amount and thus are not accessible by the measurement of the hyperfine splitting transitions. The coefficients with $j=1$ are spin-dependent anisotropic coefficients $\mathcal{T}_{w_{k10}}^\text{NR(0B)}$ and $\mathcal{T}_{w_{k10}}^\text{NR(1B)}$ that} contribute to an energy shift as given by Eq.~(48) of Ref.~\cite{PhysRevD.92.056002}.
Using the condensed notation $\Tilde{\mathcal{T}}_{w_{k10}}^\text{NR}={\mathcal{T}}_{w_{k10}}^\text{NR(0B)}+2{\mathcal{T}}_{w_{k10}}^\text{NR(1B)}$ and including the magnetic field dependence yields the following expressions for the $\sigma$ and $\pi$ transition:
\begin{equation}
    \label{eq:SME_sigma}
    \begin{aligned}
    &\nu_\sigma^{\textrm{SME}} = \nu_\sigma^{\text{SM}}  + \frac{1}{2 \pi \sqrt{3 \pi}}  \sum^{k=0,2,4}(\alpha m_r)^k(1+4\delta_{k4}) \cdot \\ & x \  (1 \! + \! x^2)^{-\frac{1}{2}} \ \left(\Tilde{\mathcal{T}}_{e_{k10}}^\text{NR,lab} -  \Tilde{\mathcal{T}}_{p_{k10}}^\text{NR,lab} \right) ,
    \end{aligned}
\end{equation}
\begin{equation}
    \label{eq:SME_pi}
    \begin{aligned}
    &\nu_\pi^{\textrm{SME}} = \nu_\pi^{\text{SM}} + \frac{1}{4 \pi \sqrt{3 \pi}}  \sum^{k=0,2,4}(\alpha m_r)^k(1+4\delta_{k4}) \cdot \\ & \left[ \Tilde{\mathcal{T}}_{e_{k10}}^\text{NR,lab}  +  \Tilde{\mathcal{T}}_{p_{k10}}^\text{NR,lab} + x \  (1 \! + \! x^2)^{-\frac{1}{2}} \left(  \Tilde{\mathcal{T}}_{e_{k10}}^\text{NR,lab} -  \Tilde{\mathcal{T}}_{p_{k10}}^\text{NR,lab} \right) \right] ,
    \end{aligned}
\end{equation}
where $x=B_{\textrm{stat}}/B_c$ with $B_c=h\nu_0/(g_+\mu_B)$ being a characteristic magnetic field of $\sim \SI{51}{\milli \tesla}$, hence $x\sim 10^{-2}$ for our magnetic field values.
Here, we have introduced $g_\pm= |g_e| \pm g_p m_e/m_p$, with ${m_e/m_p=5.446 170 214 87(33)\times10^{-4}}$ being the  ratio of the electron to proton mass, ${\mu_B=\SI[parse-numbers=false]{5.7883818060(17)\times10^{-5}}{\electronvolt\per\tesla}}$ is the Bohr magneton, ${|g_e|=2.00231930436256(35)}$ and ${g_p=5.5856946893(16)}$ are the $g$-factors of the electron and proton, respectively \cite{10.1063/5.0064853}. $\delta_{k4}$ stands for the Kronecker delta.

In the limit $x \rightarrow 0$ the $\sigma$ transition becomes insensitive to SME effects.
However, through the procedure of evaluating $\nu^c_{\pi\leftarrow\sigma}$, a suppressed SME effect on the $\sigma$ transition will be enhanced by the relative sensitivity of $\sigma$ and $\pi$ to the magnetic field (Zeeman shift) %.
given by the factor
\begin{equation}
    \label{eq:dsidpi}
    \frac{\partial \nu_\pi}{\partial \nu_\sigma} = \frac{1}{2} \left( 1 + \frac{g_-}{g_+}  \frac{\sqrt{1+x^2}}{x} \right).
\end{equation}
Since the Zeeman shift is dominated by the contribution of the electron magnetic moment, this procedure reduces the sensitivity to the SME electron coefficients by a factor $2g_p\mu_N/(g_+\mu_B)\sim0.003$, where $\mu_N=\SI[parse-numbers=false]{3.15245125844(96)\times10^{-8}}{\electronvolt\per\tesla}$ is the nuclear magneton \cite{10.1063/5.0064853}, and enhances it by a factor $2|g_e|/g_+\sim2$ for the proton coefficients:
\begin{equation}
\label{eq:deltanu_proton}
\begin{aligned}
&|h \ (\Delta\nu_\pi^+ - \Delta\nu_\pi^-)| =  \frac{\cos{\theta}}{\sqrt{3 \pi}} \sum^{k=0,2,4} (\alpha m_r)^k (1+4\delta_{k4}) \cdot \\ &
\left( \frac{2|g_e|}{g_+} \ \Tilde{\mathcal{T}}_{p_{k10}}^\text{NR,Sun}
+ \frac{2g_p\mu_N}{g_+\mu_B} \  \Tilde{\mathcal{T}}_{e_{k10}}^\text{NR,Sun} \right).
\end{aligned}
\end{equation}
 
The experimental apparatus was positioned at a latitude of $\theta_1=\SI{46.2}{\degree}$ (CERN) and the beam was oriented at an angle of $\theta_2 \approx \SI{22.5(1.5)}{\degree}$ relative to the local north, lying tangential to the surface of the Earth.
In this configuration $\cos{\theta}=  \sin{\theta_1}\sin{\theta_2} \approx 0.276(18)$. 
We hence obtain
\begin{equation}
    \label{eq:final_result}
    \mid h(\Delta \nu_{\pi}^+-\Delta\nu_{\pi}^-)\mid \frac{\sqrt{3\pi}}{\cos{\theta}} = \SI[separate-uncertainty=true]{0.9(2.3)e-21}{\giga\electronvolt} 
\end{equation}
which is consistent with zero.
The sensitivity of the measurement is used to extract the following limit on the sum of the involved spherical anisotropic SME terms, which decompose into CPT-odd ($g_{w_{k10}}^{\textrm{NR,Sun}}$) and CPT-even ($H_{w_{k10}}^{\textrm{NR,Sun}}$) coefficients: 
\begin{equation}
    \label{Constraint_all}
    \begin{aligned}
    & \Bigl| \sum^{k=0,2,4} (\alpha m_r)^k (1+4\delta_{k4}) \bigg( \frac{2|g_e|}{g_+} \Big( g_{p_{k10}}^{\textrm{NR(0B),Sun}}-H_{p_{k10}}^{\textrm{NR(0B),Sun}} \\ & + 2g_{p_{k10}}^{\textrm{NR(1B),Sun}} \! - \! 2H_{p_{k10}}^{\textrm{NR(1B),Sun}} \Big) 
    \! + \! \frac{2g_p\mu_N}{g_+\mu_B} \Big( g_{e_{k10}}^{\textrm{NR(0B),Sun}} \! - \! H_{e_{k10}}^{\textrm{NR(0B),Sun}} \\ & + 2g_{e_{k10}}^{\textrm{NR(1B),Sun}}- 2H_{e_{k10}}^{\textrm{NR(1B),Sun}} \Big) \bigg) \Bigr| < 2.3 \times 10^{-21} \, \text{GeV}.
    \end{aligned}
\end{equation}
This constitutes the first limit on this subset of non-relativistic spherical coefficients.
Assuming only one coefficient is non-zero at a time, we can extract the limits provided in Table~\ref{table:constraint:individual} on the individual coefficients.

\begin{table}[!htbp]
\centering
\begin{tabular}{||c l ||} 
 \hline
Coefficient $\mathcal{K}$& Constraint on $|\mathcal{K}|$\\ [0.5ex] 
 \hline\hline
 \multicolumn{2}{||c||}{proton} \\
$H_{p_{010}}^{\textrm{NR(0B)},\textrm{Sun}}$, $g_{p_{010}}^{\textrm{NR(0B)},\textrm{Sun}}$  & $< 1.2 \times 10^{-21}$ GeV\\[0.5ex] 
$H_{p_{010}}^{\textrm{NR(1B)},\textrm{Sun}}$, $g_{p_{010}}^{\textrm{NR(1B)},\textrm{Sun}}$  &  $< 5.8 \times 10^{-22}$ GeV\\[0.5ex] 
$H_{p_{210}}^{\textrm{NR(0B)},\textrm{Sun}}$, $g_{p_{210}}^{\textrm{NR(0B)},\textrm{Sun}}$  &  $< 8.4 \times 10^{-11}$ GeV$^{-1}$\\[0.5ex] 
$H_{p_{210}}^{\textrm{NR(1B)},\textrm{Sun}}$, $g_{p_{210}}^{\textrm{NR(1B)},\textrm{Sun}}$  & $< 4.2 \times 10^{-11}$ GeV$^{-1}$\\[0.5ex] 
$H_{p_{410}}^{\textrm{NR(0B)},\textrm{Sun}}$, $g_{p_{410}}^{\textrm{NR(0B)},\textrm{Sun}}$  & $< 1.2$ GeV$^{-3}$\\[0.5ex] 
$H_{p_{410}}^{\textrm{NR(1B)},\textrm{Sun}}$, $g_{p_{410}}^{\textrm{NR(1B)},\textrm{Sun}}$  & $< 0.6$ GeV$^{-3}$\\
 [1ex] 
  \multicolumn{2}{||c||}{electron} \\
  $H_{e_{010}}^{\textrm{NR(0B)},\textrm{Sun}}$, $g_{e_{010}}^{\textrm{NR(0B)},\textrm{Sun}}$  & $< 7.7 \times 10^{-19}$ GeV\\[0.5ex] 
$H_{e_{010}}^{\textrm{NR(1B)},\textrm{Sun}}$, $g_{e_{010}}^{\textrm{NR(1B)},\textrm{Sun}}$  &  $< 3.8 \times 10^{-19}$ GeV\\[0.5ex] 
$H_{e_{210}}^{\textrm{NR(0B)},\textrm{Sun}}$, $g_{e_{210}}^{\textrm{NR(0B)},\textrm{Sun}}$  &  $< 5.5 \times 10^{-8}$ GeV$^{-1}$\\[0.5ex] 
$H_{e_{210}}^{\textrm{NR(1B)},\textrm{Sun}}$, $g_{e_{210}}^{\textrm{NR(1B)},\textrm{Sun}}$  & $< 2.8 \times 10^{-8}$ GeV$^{-1}$\\[0.5ex] 
$H_{e_{410}}^{\textrm{NR(0B)},\textrm{Sun}}$, $g_{e_{410}}^{\textrm{NR(0B)},\textrm{Sun}}$  & $< 8.0 \times 10^{2}$ GeV$^{-3}$\\[0.5ex] 
$H_{e_{410}}^{\textrm{NR(1B)},\textrm{Sun}}$, $g_{e_{410}}^{\textrm{NR(1B)},\textrm{Sun}}$  & $< 4.0 \times 10^{2}$ GeV$^{-3}$\\
 [1ex] 
 \hline
\end{tabular}
    \caption{
    Constraints on the individual proton and electron non-relativistic spherical coefficients $\mathcal{K}^{\textrm{NR,Sun}}_{w_{k10}}$ for $k\leq4$ derived from Eq.\ref{Constraint_all} under the assumption that one coefficient is non-zero at a time.
    }
    \label{table:constraint:individual}
\end{table}

In the absence of an SME signal the ($\sigma, \pi$) pairs were used to calculate the zero-field hyperfine splitting $\nu_0^{\textrm{meas}}$ from the Breit-Rabi equations \cite{PhysRev.38.2082.2} by elimination of $B_\text{stat}$ in the equations of the transition frequencies:

\begin{equation}
    \label{eq:BreitRabi:sigma}
    \nu_\sigma(x = B_\text{stat}/B_c) = \nu_0 \sqrt{1 + x^2}, 
\end{equation}

\begin{equation}
\label{eq:BreitRabi:Pi}
    \nu_\pi(x = B_\text{stat}/B_c)= \frac{\nu_0}{2}\left(1 + \frac{g_-}{g_+} x + \sqrt{1 + x^2}\right),
\end{equation}

\begin{equation}
    \label{eq:nu0_data}
    \nu_0^\text{meas} = \frac{g_+^2(2 \nu_\pi^\text{c}-\nu_\sigma^\text{c}) + g_- \sqrt{g_-^2 (\nu_\sigma^\text{c})^2-4 g_+^2 (\nu_\pi^\text{c})^2 + 4 g_+^2 \nu_\pi^\text{c} \nu_\sigma^\text{c}}}{g_+^2 + g_-^2}.
\end{equation}
The weighted average of $\nu_0^{\textrm{meas}}$ for all pairs within a current and polarity set are shown with respect to the literature value in Fig.~\ref{fig_Results}~(c).
Analogously to the SME uncertainty treatment, correlated uncertainties have been taken into account following Ref.~\cite{Erler2015, Barlow:0471922951}.
The combined data set of 537 ($\sigma,\pi$) pairs yields $\nu^{\textrm{meas}}_0=\SI[parse-numbers=false]{1.420\,405\,751\,63(63)}{\giga\hertz}$.
The result is consistent with the literature value:
$\nu_0^{\textrm{lit}}-\nu_0^{\textrm{meas}}=\SI[parse-numbers=false]{0.14 \pm 0.59 (stat) \pm 0.23 (syst)}{\hertz}$ with a relative precision of 4.4$\times10^{-10}$ and
improves the previous determination of this quantity in a beam by a factor 6.

\section{Outlook}
\label{sec:outlook}

The ASACUSA hydrogen programme has been developed to characterise the apparatus needed for the spectroscopy of the hyperfine structure of antihydrogen in a beam, the main goal pursued by the ASACUSA-Cusp collaboration.
The work presented here has demonstrated the sub-Hz potential of the apparatus on the determination of the zero-field hyperfine transition and highlighted the intricate effects of field inhomogeneities on the resonance lineshape, which will be relevant in the antihydrogen case as well.
Measurements at opposite static magnetic field orientations lead to first constraints at the level of \SI{2.3e-21}{\giga\electronvolt} on CPT and Lorentz-violation coefficients of the SME.
Our method provides an enhanced sensitivity to proton coefficients with respect to electron coefficients by a factor equal to the ratio of the electron to proton magnetic moments.
This result is statistically limited and dominated by the determination of the magnetic field by the $\sigma$ transition.
An improvement could thus be obtained by operating at higher fields or by using advanced magnetometry methods within the cavity volume, which would remove the need for a concomitant measurement of the $\sigma$ transition.
For example, the Zeeman shift for the $\pi$ transition at the magnetic fields considered here is of the order of \SI{14}{\hertz \per \nano \tesla}.
A co-magnetometry with absolute precision of \SI{350}{\pico \tesla} would reduce the combined systematic and statistical uncertainties below \SI{5}{\hertz}, therefore improving by a factor 5 on the constraints to SME proton coefficients and more than 3 orders of magnitude on the electron coefficients.
Further improvements could be obtained with a more homogeneous static magnetic field, \cirm{as well as by using the Ramsey technique \cite{ramsey1990experiments, PhysRev.78.695, Bullis2023}}.

Precise magnetometry is also of interest in the context of antihydrogen where higher, as well as time dependent, magnetic stray fields are present in the experiment.
In the future, direct comparison between hydrogen and antihydrogen will provide constraints specifically on the CPT-odd operators.

\section*{Acknowledgements}

We wish to thank the numerous undergraduate students, CERN and SMI summer students who assisted in various ways throughout the development of this experiment.
We want to express our gratitude to the staff of the SMI Advanced Instrumentation group for hardware support.
We are indebted to the CERN cryolab team for their help, in particular to Torsten Koettig.
We would like to thank Fritz Caspers for advising on RF aspects and the TE-MSC-LMF group, in particular Attilio Milanese, Dmitry Chechenev, and Carlos Lopez for the design assistance and construction of the McKeehan coils.
We warmly thank Arnaldo Vargas for the many fruitful discussions on the Standard Model Extension.
This work was supported by the CERN-Austrian Doctoral Student Program, the Austrian Science Fund (FWF) [W1252-N27] (Doktoratskolleg Particles and Interactions), and the National Research Council of Canada (NRC).
We wish to dediacte this work to the memory of our SMI colleague Johann Zmeskal whose passion for science and unwavering kindness have left a lasting impact on all who knew him.

\end{document}